\documentclass{article}

\usepackage{amsmath}
\usepackage{MnSymbol}
\usepackage{bbm}
\usepackage{framed}
\usepackage{graphicx}
\usepackage{url}

\newcommand{\CP}{\Box}
\newcommand{\DP}{\times}
\newcommand{\SP}{\boxtimes}
\newcommand{\KS}{\oplus}
\newcommand{\KP}{\otimes}
\newcommand{\bone}{\mathbbm{1}}

\begin{document}

\title{Estimation of Laplacian spectra of direct and strong product graphs}

\author{%
Hiroki Sayama$^{1,2,3}$\\
$^1$Center for Collective Dynamics of Complex Systems\\
Binghamton University, State University of New York, Binghamton, USA\\
$^2$Max Planck Institute for the Physics of Complex Systems, Dresden, Germany\\
$^3$Center for Complex Network Research, Northeastern University, Boston, USA\\
sayama@binghamton.edu
}

\maketitle

\begin{abstract}
Calculating a product of multiple graphs has been studied in
mathematics, engineering, computer science, and more recently in
network science, particularly in the context of multilayer
networks. One of the important questions to be addressed in this area
is how to characterize spectral properties of a product graph using
those of its factor graphs. While several such characterizations have
already been obtained analytically (mostly for adjacency spectra),
characterization of Laplacian spectra of {\em direct product} and {\em
  strong product} graphs has remained an open problem. Here we develop
practical methods to estimate Laplacian spectra of direct and strong
product graphs from spectral properties of their factor graphs using a
few heuristic assumptions. Numerical experiments showed that the
proposed methods produced reasonable estimation with percentage errors
confined within a $\pm 10$\% range for most eigenvalues.
\end{abstract}

\section{Introduction}

Calculating a product of multiple graphs has been studied in several
disciplines. In mathematics, multiplication of graphs has been studied
with a particular interest in their algebraic properties as matrix
operators and their implications for topologies of resulting graphs
\cite{macduffee1933theory,sabidussi1959graph,fiedler1973algebraic,imrich1994factoring,hammack2011handbook}. Graph
products also appear in engineering as an efficient way to describe
discretized structure of objects in structural mechanics
\cite{kaveh2006block,kaveh2011laplacian}, and in computer science as a
generative model of complex networks
\cite{leskovec2005realistic,leskovec2007scalable,leskovec2010kronecker}.
More recently, graph products have also began to appear in network
science, particularly in the context of multilayer networks, where
multiplication of graphs are often used as a formal way to describe
certain types of multilayer network topologies
\cite{de2013mathematical,sole2013spectral,kivela2014multilayer,sanchez2014dimensionality,sayama2015graph}.
One of the important questions to be addressed in this area is how to
characterize spectral properties of a product graph using those of its
factor graphs, especially those of Laplacian matrices\footnote{In this
  paper, we consider simple Laplacian matrices of graphs (a.k.a.,
  combinatorial Laplacians), and not normalized Laplacians.}  because
of their high relevance to network structure and dynamics.

Several such spectral characterizations have already been obtained
analytically for certain product graphs, but they are mostly for
adjacency spectra. Characterization of Laplacian spectra has so far
been done only for {\em Cartesian product} graphs. In the meantime,
there are other important forms of graph products, such as {\em direct
  product} and {\em strong product} \cite{hammack2011handbook}, but
characterization of Laplacian spectra of those product graphs has
turned out to be quite challenging and has remained an open problem to
date.

In this paper, we attempt to address this problem by developing
practical, computationally efficient methods to estimate Laplacian
spectra of direct and strong product graphs from spectral properties
of their factor graphs, using a few heuristic assumptions. We
evaluated the effectiveness of our proposed methods through numerical
experiments, which demonstrated that they successfully produced
reasonable estimation of Laplacian spectra with percentage errors
confined within a $\pm 10$\% range for most eigenvalues.

The rest of the paper is structured as follows: In Section
\ref{definitions} we define three fundamental forms of graph products
and describe how they can be represented as operations of adjacency
matrices. In Section \ref{spectral} we summarize spectral properties
of product graphs that are already known. In Section \ref{designing}
we design our new methods to estimate Laplacian spectra of direct and
strong product graphs, and then evaluate their effectiveness by
numerical experiments. Finally, we conclude this paper with
discussions of the limitation of the current work and directions of
future research in Section \ref{conclusions}.

\section{Product graphs}
\label{definitions}

Let $G=(V_G, E_G)$ and $H=(V_H, E_H)$ be two simple connected graphs,
where $V_G$ (or $V_H$) and $E_G$ (or $E_H$) are the sets of nodes and
edges of $G$ (or $H$), respectively. We denote an adjacency matrix of
graph $X$ as $A_X$. We also use $I_n$ to represent an $n \times n$
identity matrix.

We consider operations that create a {\em product graph} of $G$ and
$H$. We call $G$ and $H$ {\em factor graphs} of the product. The node
set of a product graph will be a Cartesian product of $V_G$ and $V_H$
(i.e., $\{(g,h) \; | \; g\in V_G , \, h\in V_H\}$). Several graph
product operators have been proposed and studied in mathematics, which
differ from each other regarding how to connect those nodes in the
product graph. In this paper, we focus on the following three
fundamental graph products \cite{hammack2011handbook}:

\begin{figure}
\centering
\begin{tabular}{lcl}
$G:$ & \raisebox{-0.5\height}{\includegraphics[scale=0.32]{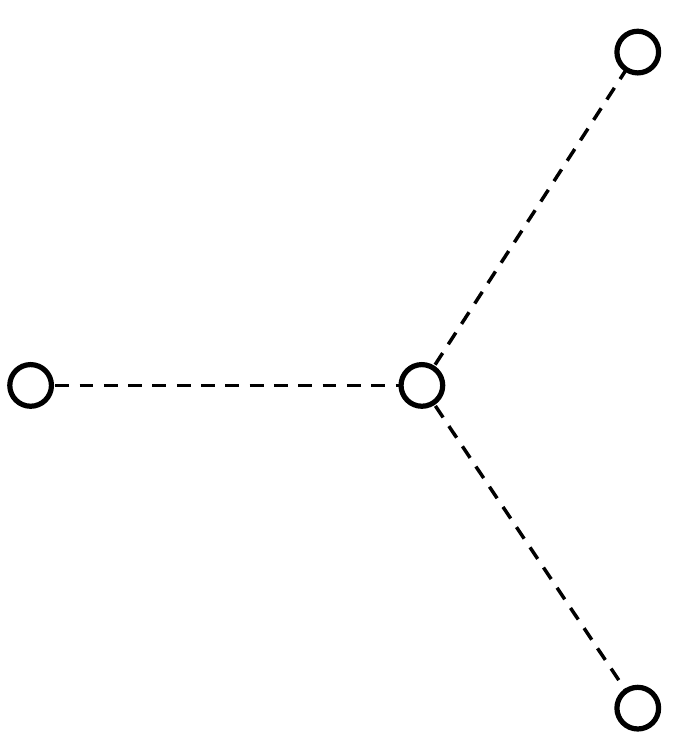}} &
$\displaystyle A_G = \begin{pmatrix}%
0 & 1 & 0 & 0\\
1 & 0 & 1 & 1\\
0 & 1 & 0 & 0\\
0 & 1 & 0 & 0
\end{pmatrix}$\\
&&\\
$H:$ & \raisebox{-0.5\height}{\includegraphics[scale=0.32]{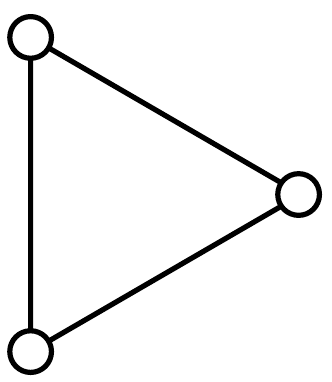}} &
$\displaystyle A_H = \begin{pmatrix}%
0 & 1 & 1\\
1 & 0 & 1\\
1 & 1 & 0
\end{pmatrix}$\\
&&\\
&&\\
\end{tabular}\\
\begin{tabular}{ll}
(a) Cartesian product $G \CP H$ & \\
\hspace*{0.6in}\raisebox{-0.5\height}{\includegraphics[scale=0.32]{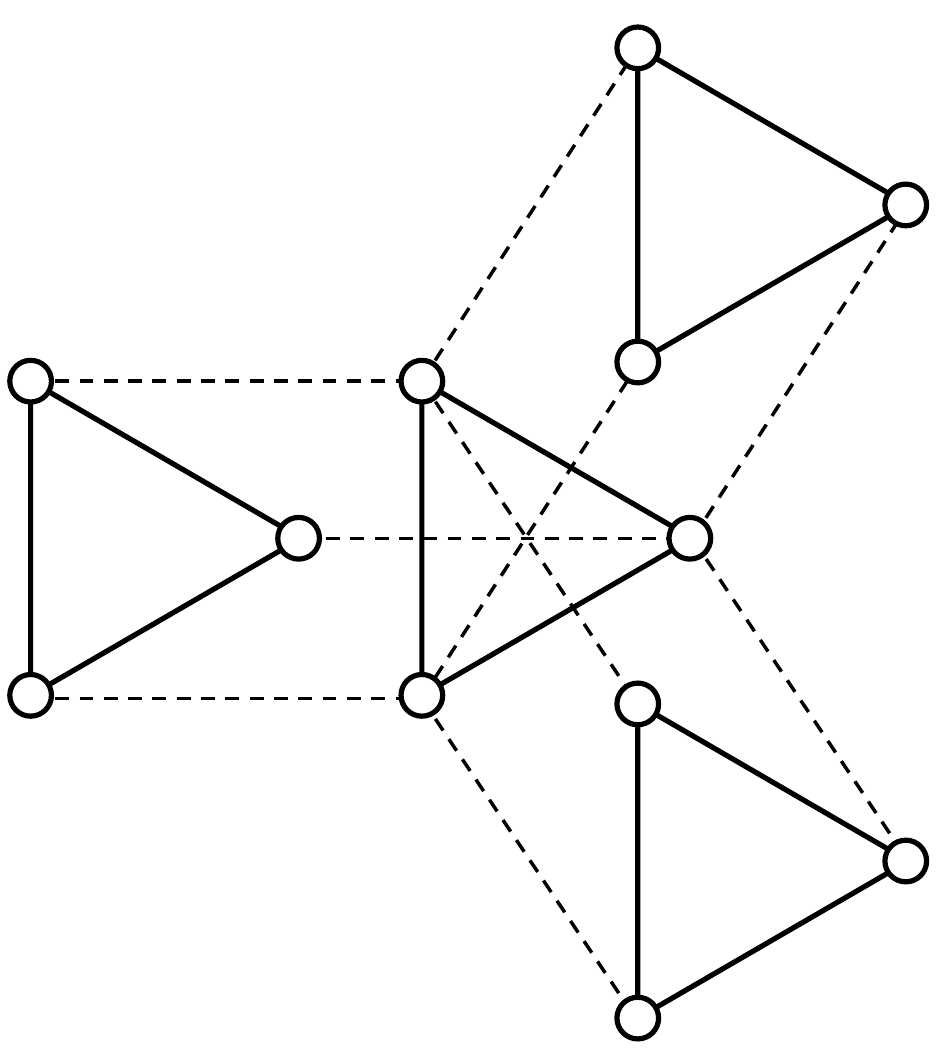}} &
\begin{minipage}{3in}
$\begin{array}{rl}%
A_{G \CP H} &= A_G \KS A_H\\
&=A_G \KP I_{|V_H|} + I_{|V_G|} \KP A_H\\
&={\footnotesize \left(\arraycolsep=2pt\def\arraystretch{0.8} \begin{array}{ccc|ccc|ccc|ccc}%
0 & 1 & 1 &     1 & 0 & 0 &     0 & 0 & 0 &     0 & 0 & 0 \\
1 & 0 & 1 &     0 & 1 & 0 &     0 & 0 & 0 &     0 & 0 & 0 \\
1 & 1 & 0 &     0 & 0 & 1 &     0 & 0 & 0 &     0 & 0 & 0 \\
\hline
1 & 0 & 0 &     0 & 1 & 1 &     1 & 0 & 0 &     1 & 0 & 0 \\
0 & 1 & 0 &     1 & 0 & 1 &     0 & 1 & 0 &     0 & 1 & 0 \\
0 & 0 & 1 &     1 & 1 & 0 &     0 & 0 & 1 &     0 & 0 & 1 \\
\hline
0 & 0 & 0 &     1 & 0 & 0 &     0 & 1 & 1 &     0 & 0 & 0 \\
0 & 0 & 0 &     0 & 1 & 0 &     1 & 0 & 1 &     0 & 0 & 0 \\
0 & 0 & 0 &     0 & 0 & 1 &     1 & 1 & 0 &     0 & 0 & 0 \\
\hline
0 & 0 & 0 &     1 & 0 & 0 &     0 & 0 & 0 &     0 & 1 & 1 \\
0 & 0 & 0 &     0 & 1 & 0 &     0 & 0 & 0 &     1 & 0 & 1 \\
0 & 0 & 0 &     0 & 0 & 1 &     0 & 0 & 0 &     1 & 1 & 0
\end{array}\right)}
\end{array}$
\end{minipage}\\
&\\
(b) Direct (tensor) product $G \DP H$ & \\
\hspace*{0.6in}\raisebox{-0.5\height}{\includegraphics[scale=0.32]{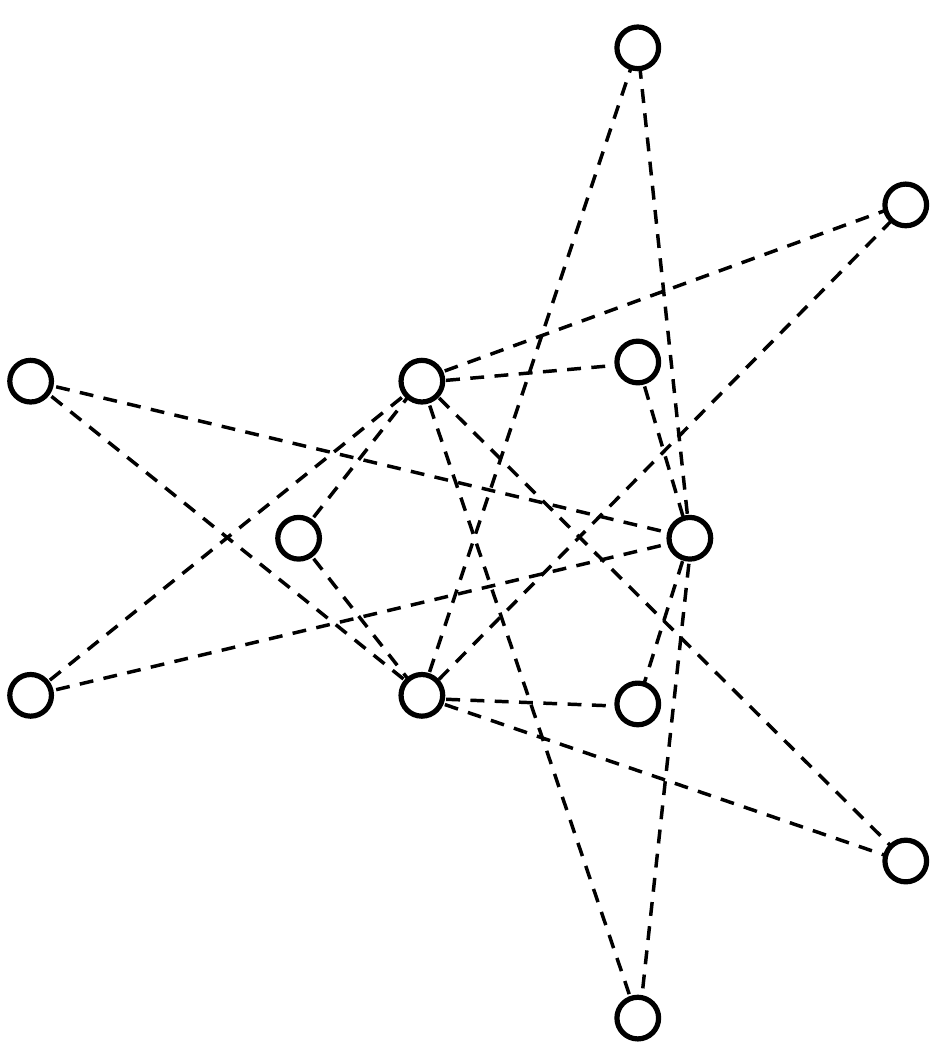}} &
\begin{minipage}{3in}
$\begin{array}{rl}%
A_{G \DP H} &= A_G \KP A_H\\
&={\footnotesize \left(\arraycolsep=2pt\def\arraystretch{0.8} \begin{array}{ccc|ccc|ccc|ccc}%
0 & 0 & 0 &     0 & 1 & 1 &     0 & 0 & 0 &     0 & 0 & 0 \\
0 & 0 & 0 &     1 & 0 & 1 &     0 & 0 & 0 &     0 & 0 & 0 \\
0 & 0 & 0 &     1 & 1 & 0 &     0 & 0 & 0 &     0 & 0 & 0 \\
\hline
0 & 1 & 1 &     0 & 0 & 0 &     0 & 1 & 1 &     0 & 1 & 1 \\
1 & 0 & 1 &     0 & 0 & 0 &     1 & 0 & 1 &     1 & 0 & 1 \\
1 & 1 & 0 &     0 & 0 & 0 &     1 & 1 & 0 &     1 & 1 & 0 \\
\hline
0 & 0 & 0 &     0 & 1 & 1 &     0 & 0 & 0 &     0 & 0 & 0 \\
0 & 0 & 0 &     1 & 0 & 1 &     0 & 0 & 0 &     0 & 0 & 0 \\
0 & 0 & 0 &     1 & 1 & 0 &     0 & 0 & 0 &     0 & 0 & 0 \\
\hline
0 & 0 & 0 &     0 & 1 & 1 &     0 & 0 & 0 &     0 & 0 & 0 \\
0 & 0 & 0 &     1 & 0 & 1 &     0 & 0 & 0 &     0 & 0 & 0 \\
0 & 0 & 0 &     1 & 1 & 0 &     0 & 0 & 0 &     0 & 0 & 0
\end{array}\right)}
\end{array}$
\end{minipage}\\
&\\
(c) Strong product $G \SP H$ & \\
\hspace*{0.6in}\raisebox{-0.5\height}{\includegraphics[scale=0.32]{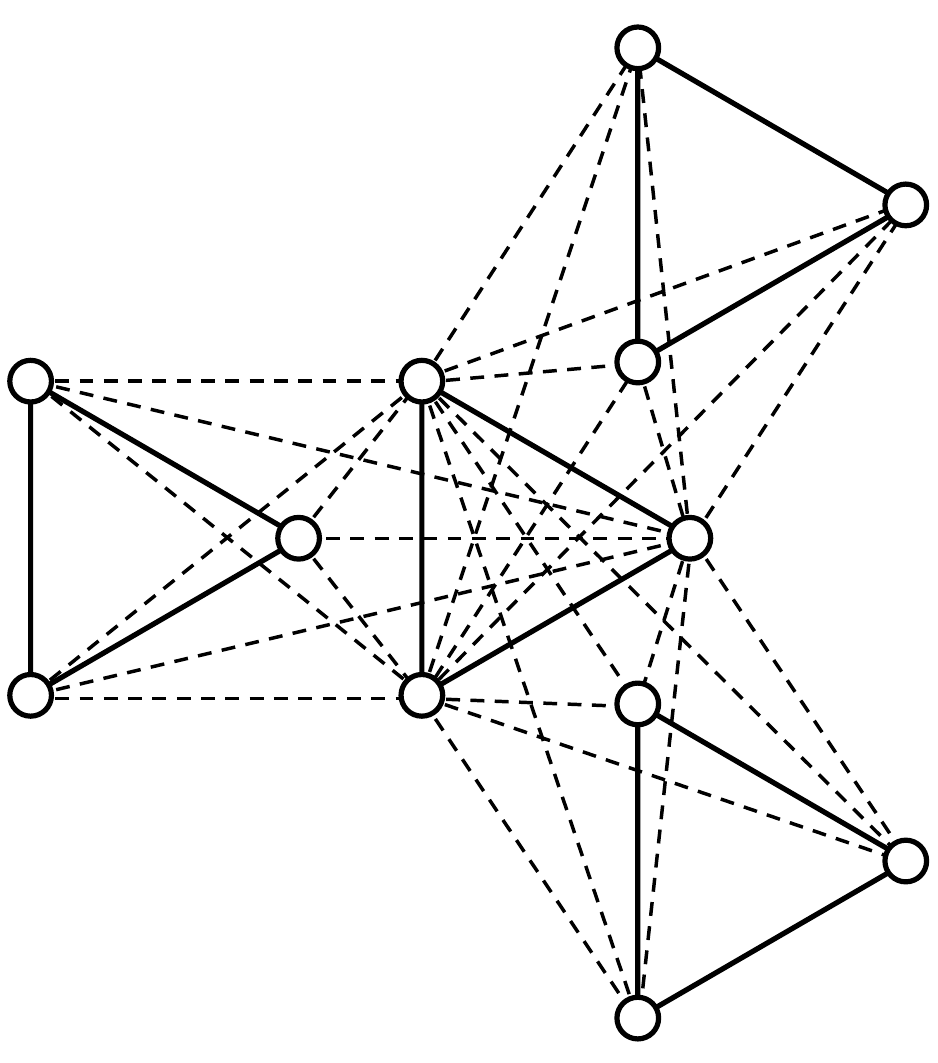}} &
\begin{minipage}{3in}
$\begin{array}{rl}%
A_{G \SP H} &= A_G \KS A_H + A_G \KP A_H\\
&= (A_G + I_{|V_G|}) \KP (A_H + I_{|V_H|}) - I_{|V_G|} \KP I_{|V_H|}\\
&={\footnotesize \left(\arraycolsep=2pt\def\arraystretch{0.8} \begin{array}{ccc|ccc|ccc|ccc}%
0 & 1 & 1 &     1 & 1 & 1 &     0 & 0 & 0 &     0 & 0 & 0 \\
1 & 0 & 1 &     1 & 1 & 1 &     0 & 0 & 0 &     0 & 0 & 0 \\
1 & 1 & 0 &     1 & 1 & 1 &     0 & 0 & 0 &     0 & 0 & 0 \\
\hline
1 & 1 & 1 &     0 & 1 & 1 &     1 & 1 & 1 &     1 & 1 & 1 \\
1 & 1 & 1 &     1 & 0 & 1 &     1 & 1 & 1 &     1 & 1 & 1 \\
1 & 1 & 1 &     1 & 1 & 0 &     1 & 1 & 1 &     1 & 1 & 1 \\
\hline
0 & 0 & 0 &     1 & 1 & 1 &     0 & 1 & 1 &     0 & 0 & 0 \\
0 & 0 & 0 &     1 & 1 & 1 &     1 & 0 & 1 &     0 & 0 & 0 \\
0 & 0 & 0 &     1 & 1 & 1 &     1 & 1 & 0 &     0 & 0 & 0 \\
\hline
0 & 0 & 0 &     1 & 1 & 1 &     0 & 0 & 0 &     0 & 1 & 1 \\
0 & 0 & 0 &     1 & 1 & 1 &     0 & 0 & 0 &     1 & 0 & 1 \\
0 & 0 & 0 &     1 & 1 & 1 &     0 & 0 & 0 &     1 & 1 & 0
\end{array}\right)}
\end{array}$
\end{minipage}
\end{tabular}
\caption{Three fundamental graph products. Top: Two factor graphs used
  in this example, $G$ and $H$, and their adjacency matrices. (a)
  Cartesian product. (b) Direct (tensor) product. (c) Strong
  product. Solid and dashed lines are intra-layer and inter-layer
  edges, respectively. Here, $G$ is considered an inter-layer network
  that connects the intra-layer network $H$, but one can consider the
  other way around too.}
\label{fig:graph-products}
\end{figure}

\begin{description}

\item[{\em Cartesian product}:] Denoted as $G \CP H$. Two nodes
  $(g,h)$ and $(g',h')$ are connected in $G \CP H$ if and only if
\begin{equation}
g = g' \text{~~and~~} (h,h') \in E_H , \quad \text{or} \quad
(g,g') \in E_G \text{~~and~~} h = h' .
\end{equation}
The adjacency matrix of $G \CP H$ is given by
\begin{align}
A_{G \CP H} &= A_G \KS A_H\\
&=A_G \KP I_{|V_H|} + I_{|V_G|} \KP A_H , \label{eq:CP-def}
\end{align}
where $\KS$ and $\KP$ denote a Kronecker sum
and a Kronecker product of matrices, respectively. An example is shown
in Fig.~\ref{fig:graph-products}(a).

\item[{\em Direct (tensor) product}:] Denoted as $G \DP H$. Two nodes
  $(g,h)$ and $(g',h')$ are connected in $G \DP H$ if and only if
\begin{equation}
(g,g') \in E_G \text{~~and~~} (h,h') \in E_H .
\end{equation}
The adjacency matrix of $G \DP H$ is given by
\begin{align}
A_{G \DP H} &= A_G \KP A_H .
\end{align}
An example is shown in Fig.~\ref{fig:graph-products}(b).

\item[{\em Strong product}:] Denoted as $G \SP H$. Two nodes
  $(g,h)$ and $(g',h')$ are connected in $G \SP H$ if and only if
\begin{equation}
g = g' \text{~~and~~} (h,h') \in E_H , \quad \text{or} \quad
(g,g') \in E_G \text{~~and~~} h = h' , \quad \text{or} \quad
(g,g') \in E_G \text{~~and~~} (h,h') \in E_H .
\end{equation}
The adjacency matrix of $G \SP H$ is given by
\begin{align}
A_{G \SP H} &= A_G \KS A_H + A_G \KP A_H\\
&= (A_G + I_{|V_G|}) \KP (A_H + I_{|V_H|}) - I_{|V_G|} \KP I_{|V_H|} .
\end{align}
As seen above, a strong product is a sum of Cartesian and direct
products. An example is shown in Fig.~\ref{fig:graph-products}(c).

\end{description}
All of these three graph products are commutative, in the sense that
$G * H$ and $H * G$ (where $*$ can be either $\CP$, $\DP$, or $\SP$)
are isomorphic to each other\footnote{The resulting adjacency matrices
  will be different, but there is always a permutation of rows/columns
  to make them identical.}. These operations are also associative.

\section{Spectral properties of product graphs}
\label{spectral}

Relationships between spectral properties of a product graph and those
of its factor graphs have been known for degree and adjacency spectra
for all of the three products, as well as Laplacian spectra for
Cartesian product
\cite{macduffee1933theory,fiedler1973algebraic,kaveh2011laplacian,barik2015laplacian}. They
are summarized below.

\subsection{Degree spectra}

Let $\bone_n$ be an all-one column vector with length $n$. Then the
degree spectrum (i.e., degree sequence) of graph $X$ can be obtained
as $d^X = A_X \bone_{|V_X|}$. Here we denote the degree spectra of
graphs $G$ and $H$ as $d^G=\big(d^G_i\big)$ and $d^H=\big(d^H_j\big)$,
respectively ($i = 1, 2, \ldots, |V_G|, \; j = 1, 2, \ldots,
|V_H|$). Applying this to the adjacency matrices of product graphs
yields the following (note that $\bone_{|V_G|\cdot|V_H|} =
\bone_{|V_G|} \KP \bone_{|V_H|}$):

~\\\noindent Cartesian product:
\begin{align}
A_{G \CP H} \bone_{|V_G|\cdot|V_H|}
&=(A_G \KP I_{|V_H|}) (\bone_{|V_G|} \KP \bone_{|V_H|}) + (I_{|V_G|} \KP A_H) (\bone_{|V_G|} \KP \bone_{|V_H|})\\
&=d^G \KP \bone_{|V_H|} + \bone_{|V_G|} \KP d^H\\
&=\bigg(d^G_i + d^H_j\bigg)
\end{align}
Direct product:
\begin{align}
A_{G \DP H} \bone_{|V_G|\cdot|V_H|}
&=(A_G \KP A_H) (\bone_{|V_G|} \KP \bone_{|V_H|})\\
&=d^G \KP d^H\\
&=\bigg(d^G_i \, d^H_j\bigg)
\end{align}
Strong product:
\begin{align}
A_{G \SP H} \bone_{|V_G|\cdot|V_H|}
&= A_{G \CP H} \bone_{|V_G|\cdot|V_H|} + A_{G \DP H} \bone_{|V_G|\cdot|V_H|}\\
&=\bigg(d^G_i + d^H_j + d^G_i \, d^H_j\bigg)
\end{align}

\subsection{Adjacency spectra}

Adjacency spectra (i.e., eigenvalues of adjacency matrices) of product
graphs can be obtained in a similar manner. Let $\{\lambda^G_i\}$ and
$\{\lambda^H_j\}$ be the eigenvalues of $A_G$ and $A_H$ with
corresponding eigenvectors $\{v^G_i\}$ and $\{v^H_j\}$, respectively
($i = 1, 2, \ldots, |V_G|, \; j = 1, 2, \ldots, |V_H|$). Then it can
be shown that for all of the three product graphs, their adjacency
matrices have $v^G_i \KP v^H_j$ as eigenvectors, as follows:

~\\\noindent Cartesian product:
\begin{align}
A_{G \CP H} (v^G_i \KP v^H_j)
&=(A_G \KP I_{|V_H|}) (v^G_i \KP v^H_j) + (I_{|V_G|} \KP A_H) (v^G_i \KP v^H_j)\\
&=\lambda^G_i v^G_i \KP v^H_j + v^G_i \KP \lambda^H_j v^H_j\\
&=\bigg(\lambda^G_i + \lambda^H_j\bigg) (v^G_i \KP v^H_j) \label{eq:CP-sprel}
\end{align}
Direct product:
\begin{align}
A_{G \DP H} (v^G_i \KP v^H_j)
&=(A_G \KP A_H) (v^G_i \KP v^H_j)\\
&=\lambda^G_i v^G_i \KP \lambda^H_j v^H_j\\
&=\bigg(\lambda^G_i \, \lambda^H_j\bigg) (v^G_i \KP v^H_j)
\end{align}
Strong product:
\begin{align}
A_{G \SP H} (v^G_i \KP v^H_j)
&= A_{G \CP H} (v^G_i \KP v^H_j) + A_{G \DP H} (v^G_i \KP v^H_j)\\
&=\bigg(\lambda^G_i + \lambda^H_j + \lambda^G_i \, \lambda^H_j \bigg) (v^G_i \KP v^H_j)
\end{align}
Note the similarity of results between degree and adjacency spectra.

\subsection{Laplacian spectra for Cartesian product graphs}

Laplacian spectra of product graphs turn out to be not as simply
characterizable as the other two shown above. So far, only a
Laplacian spectrum of a Cartesian product graph has been analytically
linked to Laplacian spectra of its factor graphs. Here we denote the
Laplacian and degree matrices of graph $X$ as $L_X$ and $D_X$,
respectively. Then the Laplacian matrix of Cartesian product graph $G
\CP H$ can be characterized as follows:
\begin{align}
L_{G \CP H} &= D_{G \CP H} - A_{G \CP H}\\
&=(D_G \KP I_{|V_H} + I_{|V_G|} \KP D_H) - (A_G \KP I_{|V_H|} + I_{|V_G|} \KP A_H)\\
&=D_G \KP I_{|V_H} + I_{|V_G|} \KP D_H - (D_G - L_G) \KP I_{|V_H|} - I_{|V_G|} \KP (D_H - L_H)\\
&=L_G \KP I_{|V_H|} + I_{|V_G|} \KP L_H
\end{align}
The result above shows that the relationship between Laplacians of
factor and product graphs are identical to the relationship between
their adjacency matrices (Eq.~(\ref{eq:CP-def})). Therefore, the same
conclusion about their spectral relationship naturally follows. Let
$\{\mu^G_i\}$ and $\{\mu^H_j\}$ be the eigenvalues of $L_G$ and $L_H$
with corresponding eigenvectors $\{w^G_i\}$ and $\{w^H_j\}$,
respectively ($i = 1, 2, \ldots, |V_G|, \; j = 1, 2, \ldots,
|V_H|$). Then the following can be shown directly from
Eq.~(\ref{eq:CP-sprel}):
\begin{align}
L_{G \CP H} (w^G_i \KP w^H_j) &=\bigg(\mu^G_i + \mu^H_j\bigg) (w^G_i
\KP w^H_j)
\end{align}

\section{Laplacian spectra of direct and strong product graphs}
\label{designing}

Characterizing Laplacian spectra of direct product and strong product
graphs is quite challenging and has remained an open problem. The
objective of the present study is to develop practical methods to
estimate their Laplacian spectra using a few heuristic assumptions, in
view of potential applications for large-scale multilayer network
analysis
\cite{sanchez2014dimensionality,kivela2014multilayer,boccaletti2014structure}.

\subsection{Estimating Laplacian spectra of direct product graphs}

\begin{figure}[tbp]
\centering
\includegraphics[width=0.6\textwidth]{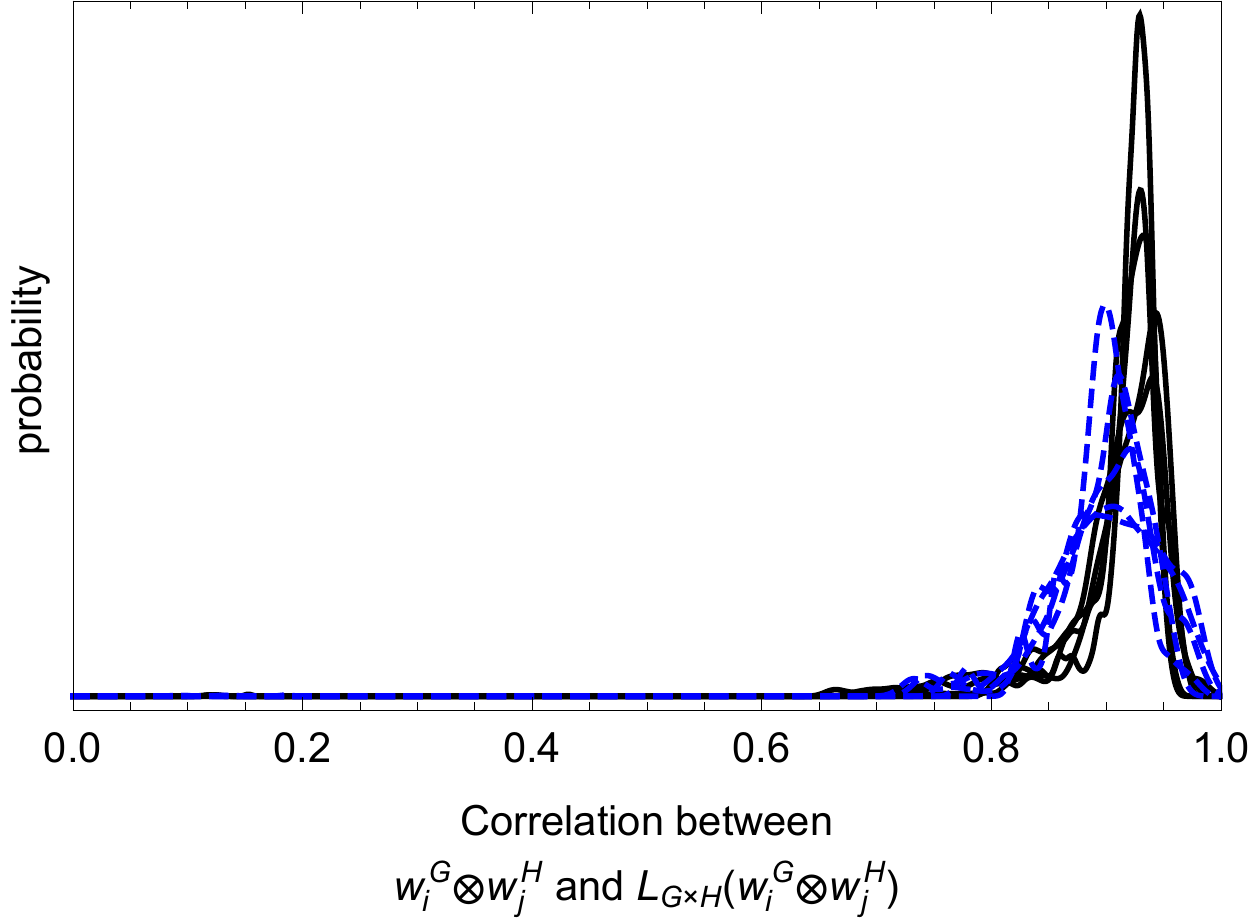}
\caption{Smoothed probability density functions of vector correlation
  coefficients between $w^G_i \KP w^H_j$ and $L_{G \DP H} (w^G_i \KP
  w^H_j)$. Each curve is numerically obtained from a direct product of
  two randomly generated graphs, $G$ and $H$, by measuring the
  distribution of vector correlation coefficients between $w^G_i \KP
  w^H_j$ and $L_{G \DP H} (w^G_i \KP w^H_j)$ over all possible
  combinations of $i$ and $j$, where $w^G_i$ and $w^H_j$ are the
  eigenvectors of $L_G$ and $L_H$, respectively. Five independent
  numerical results are drawn for each of the following two
  conditions. Black (solid) curves: $G$ = Erd\H os-R\'enyi random
  graph with 50 nodes and 100 edges, and $H$ = Erd\H os-R\'enyi random
  graph with 30 nodes and 90 edges. Blue (dashed) curves: $G$ =
  Barab\'asi-Albert graph with 50 nodes and 2 edges per each newcomer
  node, and $H$ = Barab\'asi-Albert graph with 30 nodes and 3 edges
  per each newcomer node. These conditions are also used in numerical
  experiments described later in this paper. As seen in this figure,
  the correlation coefficients are above 0.8 most of the cases. This
  observation remains similar when the sizes of the graphs are varied
  (the correlations are stronger for larger graphs).}
\label{fig:correlations}
\end{figure}

A Laplacian of direct product graph $G \DP H$ is given as follows:
\begin{align}
L_{G \DP H} &= D_{G \DP H} - A_{G \DP H}\\
&=(D_G \KP D_H) - (A_G \KP A_H)\\
&=D_G \KP D_H - (D_G - L_G) \KP (D_H - L_H)\\
&= L_G \KP D_H + D_G \KP L_H - L_G \KP L_H \label{eq:LGDPH}
\end{align}
A general solution for obtaining eigenvalues and eigenvectors of this
Laplacian matrix from those of its factor graphs is not known to
date. The complexity of the problem partly comes from the involvement
of $D_G$ and $D_H$ in the formula above, which were successfully
eliminated in all of the previous cases. (We will come back to this
point later.)  In the meantime, however, there are partial
regularities known for Laplacian spectra of direct product graphs that
resemble those of Cartesian product graphs, especially when either
factor graph is regular \cite{barik2015laplacian}. Moreover, there is
empirical evidence that $w^G_i \KP w^H_j$, i.e., eigenvectors of $L_{G
  \CP H}$, are relatively close to eigenvectors of $L_{G \DP H}$,
i.e.,
\begin{align}
L_{G \DP H}(w^G_i \KP w^H_j) \approx \alpha (w^G_i \KP w^H_j),
\end{align}
which can be numerically observed by measuring vector correlations
between $w^G_i \KP w^H_j$ and $L_{G \DP H} (w^G_i \KP w^H_j)$
(Fig.~\ref{fig:correlations}). These clues lead us to make an
assumption that $w^G_i \KP w^H_j$ could be used as a reasonable
substitute of the true eigenvectors of $L_{G \DP H}$, at least to
facilitate the process of estimating its spectrum.

Using the assumption made above, we attempt to calculate $L_{G \DP H}
(w^G_i \KP w^H_j)$ as follows:
\begin{align}
L_{G \DP H} (w^G_i \KP w^H_j) &= (L_G \KP D_H + D_G \KP L_H - L_G \KP L_H)(w^G_i \KP w^H_j)\\
&= (L_G \KP D_H)(w^G_i \KP w^H_j) + (D_G \KP L_H)(w^G_i \KP w^H_j) - (L_G \KP L_H)(w^G_i \KP w^H_j)\\
&= \mu^G_i w^G_i \KP D_H w^H_j + D_G w^G_i \KP \mu^H_j w^H_j - \mu^G_i w^G_i \KP \mu^H_j w^H_j \label{eq:LGDPH2}
\end{align}
This result can also be written in an aggregate form for all
eigenvalues and eigenvectors. Let $W_G$ and $W_H$ be
$|V_G|\times|V_G|$ and $|V_H|\times|V_H|$ square matrices that contain
all $w^G_i$ and $w^H_j$ as column vectors, respectively. Also, let
$M_G$ and $M_H$ be diagonal matrices whose diagonal entries are
$\mu^G_i$ and $\mu^H_j$ in the same orders as $w^G_i$ in $W_G$ and
$w^H_j$ in $W_H$, respectively. Using these matrices,
Eq.~(\ref{eq:LGDPH2}) can be rewritten as
\begin{align}
L_{G \DP H} (W_G \KP W_H) &= (L_G \KP D_H + D_G \KP L_H - L_G \KP L_H)(W_G \KP W_H)\\
&= W_G M_G \KP D_H W_H + D_G W_G \KP W_H M_H - W_G M_G \KP W_H M_H.
\end{align}
At this stage, we are no longer able to simplify the result any
further, because we used $w^G_i \KP w^H_j$ as hypothetical
eigenvectors that are actually not. However, if we could make another
bold (mathematically incorrect) assumption that we could let $D_G W_G
\approx W_G D_G$ and $D_H W_H \approx W_H D_H$, the above formula could be
decomposed further as
\begin{align}
L_{G \DP H} (W_G \KP W_H)
&\approx W_G M_G \KP W_H D_H + W_G D_G \KP W_H M_H - W_G M_G \KP W_H M_H\\
&= (W_G \KP W_H) \bigg(M_G \KP D_H + D_G \KP M_H - M_G \KP M_H \bigg) ,
\end{align}
where inside the second pair of parentheses is a diagonal matrix whose
diagonal entries show a hypothetical spectrum:
\begin{align}
&\{ \; \mu^G_i d^H_j + d^G_i \mu^H_j - \mu^G_i \mu^H_j \; \} \label{eq:estimated-spectrum}
\end{align}
Needless to say, this is not a valid conclusion because we used two
mathematically incorrect assumptions. Yet this result has an
intriguing symmetry with Eq.~(\ref{eq:LGDPH}) and it also perfectly
agrees with partial properties of eigenvalues of $L_{G \DP H}$
reported elsewhere \cite{barik2015laplacian}.

The main question is now whether the heuristic assumptions made above,
\begin{align}
&L_{G \DP H}(w^G_i \KP w^H_j) \approx \alpha (w^G_i \KP w^H_j), \quad \text{and}\\
&D_G W_G \approx W_G D_G \quad \text{and} \quad D_H W_H \approx W_H D_H,
\end{align}
can be reasonable substitutions or not. These become exact equalities
if $G$ and $H$ are regular graphs (i.e., if $D_G$ and $D_H$ are scalar
multiplications of identity matrices), which is not the case with
non-homogeneous degree spectra.

Here we note, however, an important fact that the orderings of $w^G_i$
and $w^H_j$ in $W_G$ and $W_H$ (and hence $\mu^G_i$ and $\mu^H_j$ in
$M_G$ and $M_H$) are independent of node orderings in $D_G$ and $D_H$,
respectively. This means that one could reduce the mathematical
inaccuracy arising from these two incorrect assumptions by finding
optimal column permutations of $W_G$ and $W_H$ (which also apply to
$M_G$ and $M_H$). In some sense, the involvement of $D_G$ and $D_H$,
which was the primary source of complexity of the problem, also brings
an opportunity we could exploit to mitigate the damage caused by our
sloppy mathematical derivation, which we will definitely try in what
follows.

One thing that is immediately apparent is that it would be impractical
to try to find true optimal orderings. This is firstly because the
search space of this optimization problem grows combinatorially as the
size of $G$ and $H$increases, and secondly because the full sets of
eigenvectors ($W_G$ and $W_H$) would not be available in a realistic
scenario of large-scale network analysis. We therefore tested the
following five easily implementable heuristic methods that use only
degrees and eigenvalues of factor graphs. In each method, it is
assumed that the degree sequences ($d^G_i$ and $d^H_j$) are already
sorted in an ascending order, while the orders of eigenvalues
($\mu^G_i$ and $\mu^H_j$) are altered differently:
\begin{enumerate}
\item {\em Uncorrelated ordering:} $\mu^G_i$ and $\mu^H_j$ are
  randomly permuted.
\item {\em Correlated ordering:} $\mu^G_i$ and $\mu^H_j$ are sorted in
  ascending order, naturally inducing positive correlations with
  $d^G_i$ and $d^H_j$, respectively.
\item {\em Correlated ordering with randomization:} Same as above,
  except that each value of $\mu^G_i$ and $\mu^H_j$ is multiplied by a
  random number sampled from a uniform distribution $[0.9, 1.1]$
  (i.e., the value is randomly varied within a $\pm 10$\% range) before
  sorting.
\item {\em Anti-correlated ordering:} $\mu^G_i$ and $\mu^H_j$ are
  sorted in descending order, naturally inducing negative
  correlations with $d^G_i$ and $d^H_j$, respectively.
\item {\em Anti-correlated ordering with randomization:} Same as
  above, except that each value of $\mu^G_i$ and $\mu^H_j$ is
  multiplied by a random number sampled from a uniform distribution
  $[0.9, 1.1]$ (i.e., the value is randomly varied within a $\pm 10$\%
  range) before sorting.
\end{enumerate}
Methods 3 and 5 were included in the above list to represent
intermediate cases between completely random and completely sorted
methods.

To compare the performance of these methods, we applied each of them
to the same types of networks as used in
Fig.~\ref{fig:correlations}. For each combination of two factor graphs
and a method, we calculated the actual spectrum of $L_{G \DP H}$
numerically, as well as the estimated spectrum using
Eq.~(\ref{eq:estimated-spectrum}) with specific orderings of $\mu^G_i$
and $\mu^H_j$ determined by each ordering method. Both the actual and
estimated spectra were sorted, and then each pair of actual and
estimated eigenvalues were compared and their RMSE (root mean square
error) was calculated across all eigenvalues, as the overall
performance measure of each method.

\begin{figure}[tbp]
\centering
\includegraphics[width=\textwidth]{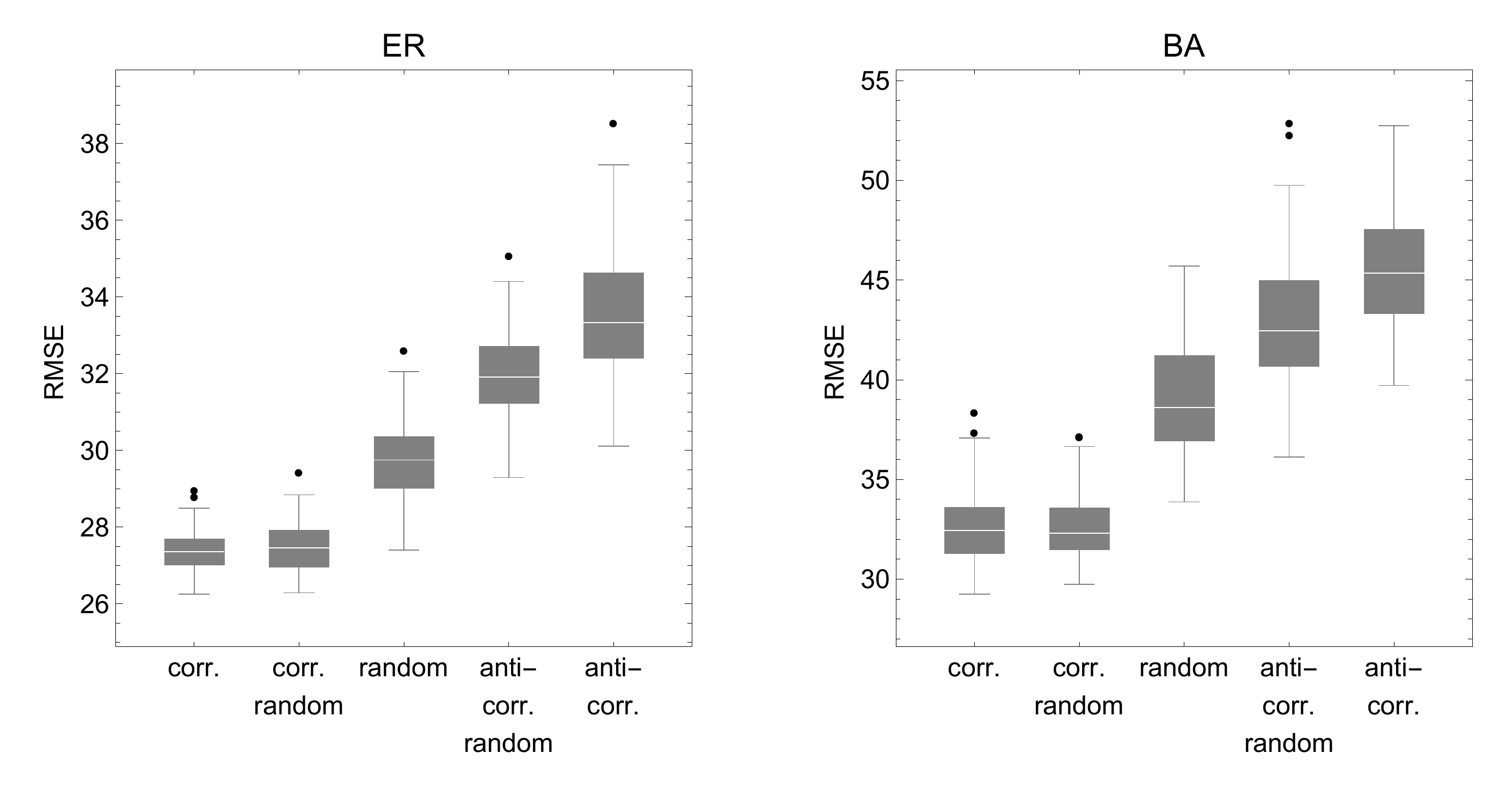}
\caption{Comparison of estimation performance (RMSE) among five
  ordering methods (see text for details of each method) for
  estimating Laplacian spectra of direct product graphs. Results were
  collected from one hundred independent tests for each condition, and
  their distributions were shown in box-whisker plots. Gray boxes show
  25 percentiles above and below a mean, while whiskers show the whole
  range. Black dots indicate outliers. Left (``ER''): $G$ = Erd\H
  os-R\'enyi random graph with 50 nodes and 100 edges, and $H$ = Erd\H
  os-R\'enyi random graph with 30 nodes and 90 edges. Right (``BA''):
  $G$ = Barab\'asi-Albert graph with 50 nodes and 2 edges per each
  newcomer node, and $H$ = Barab\'asi-Albert graph with 30 nodes and 3
  edges per each newcomer node. ANOVA and Tukey/Bonferroni posthoc
  tests showed highly significant differences among the conditions
  except between correlated ordering (``corr.'')  and correlated
  ordering with randomization (``corr. random'').}
\label{fig:performance-evaluation}
\end{figure}

Results are summarized in Fig.~\ref{fig:performance-evaluation}, which
shows a clear trend for both Erd\H os-R\'enyi and Barab\'asi-Albert
factor graphs. The most effective ordering methods turned out to be
correlated orderings (without and with randomization; there was no
statistical difference between them). In contrast, anti-correlated
orderings had an adverse effect on the estimation results. Based on
this result, we adopted the correlated ordering, the simplest and best
performing choice among the five methods tested.

Our final estimation method is summarized as follows:
\begin{framed}
\noindent {\bf Proposed method for estimating a Laplacian spectrum of
  a direct product graph}
\begin{enumerate}
\item Obtain degree and Laplacian spectra of two factor graphs $G$ and
  $H$. We denote them as $d^G=\{d^G_i \}$, $d^H=\{d^H_j\}$,
  $\mu^G=\{\mu^G_i\}$ and $\mu^H=\{\mu^H_j\}$.
\item Sort all the spectra in an ascending order.
\item Calculate the following set for all $i = 1, 2, \ldots, |V_G|$
  and $j = 1, 2, \ldots, |V_H|$:
\[ \mu^{G \DP H} = \{ \; \mu^G_i d^H_j + d^G_i \mu^H_j - \mu^G_i \mu^H_j \; \} \]
$\mu^{G \DP H}$ is the estimated Laplacian spectrum of $G \DP H$.
\end{enumerate}
\end{framed}
The computational complexity of this method is $O(f(m) + f(n) + m \log
m + n \log n + mn)$, where $m = |V_G|$, $n = |V_H|$, and $O(f(k))$ is
the computational complexity of calculating degree and Laplacian
spectra of a graph made of $k$ nodes. In general $O(f(k)) = O(k^3)$,
so the complexity of this method is characterized by $O(m^3 + n^3 + m
\log m + n \log n + mn)$. This is substantially smaller than $O(f(mn))
= O(m^3n^3)$, i.e., the complexity of explicit computation of
eigenvalues of $L_{G \DP H}$, especially when $m$ and $n$ are large.

\begin{figure}[tbp]
\centering
\includegraphics[width=0.5\textwidth]{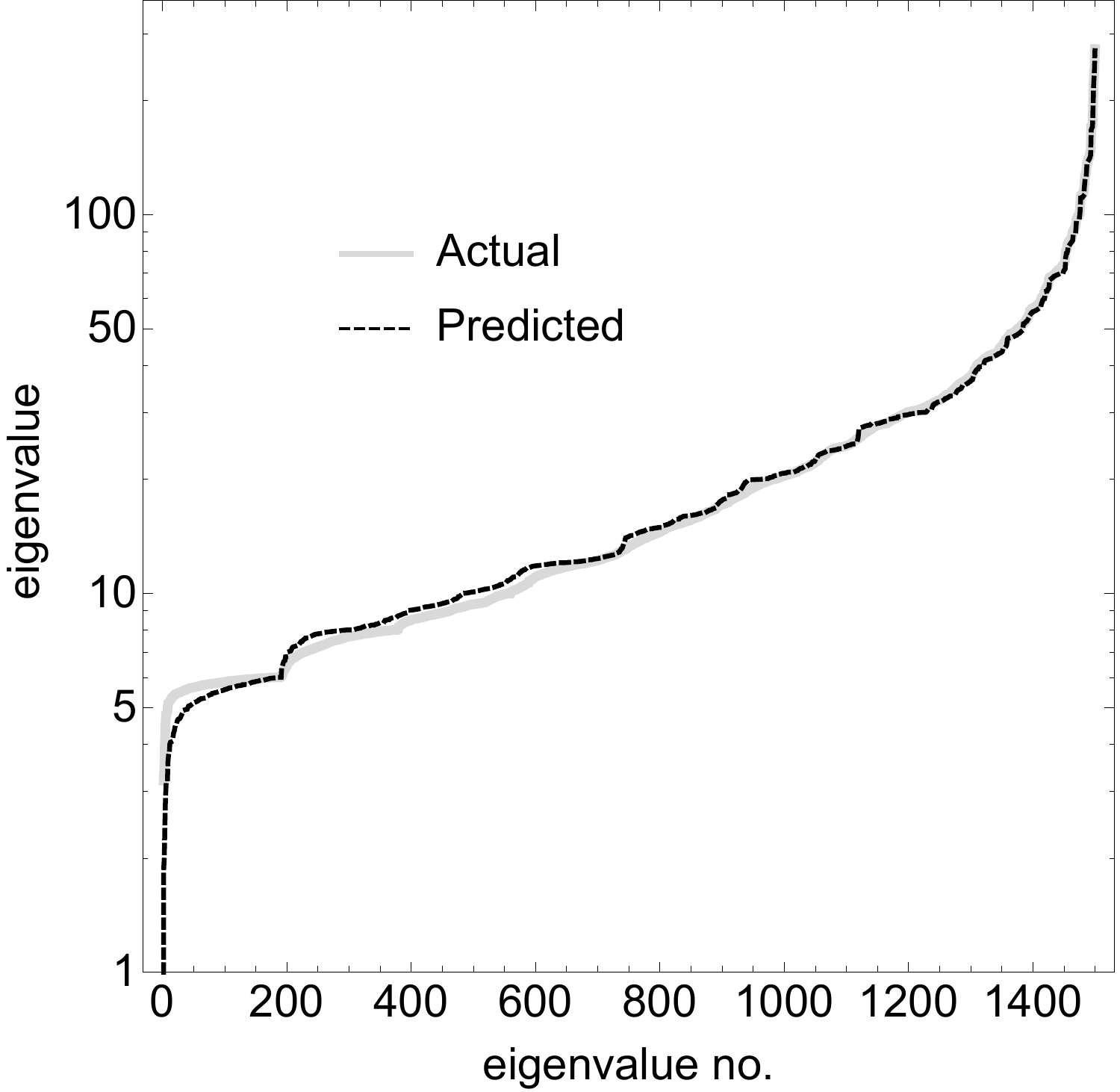}
\caption{An example of a Laplacian spectrum of a direct product graph
  estimated using the proposed method (black dashed curve) in
  comparison with the actual one (gray solid curve). Eigenvalues are
  sorted in an ascending order. $G$ = Barab\'asi-Albert graph with 50
  nodes and 2 edges per each newcomer node, and $H$ =
  Barab\'asi-Albert graph with 30 nodes and 3 edges per each newcomer
  node.}
\label{fig:prediction-example}
\end{figure}

\begin{figure}[tbp]
\centering
\includegraphics[width=\textwidth]{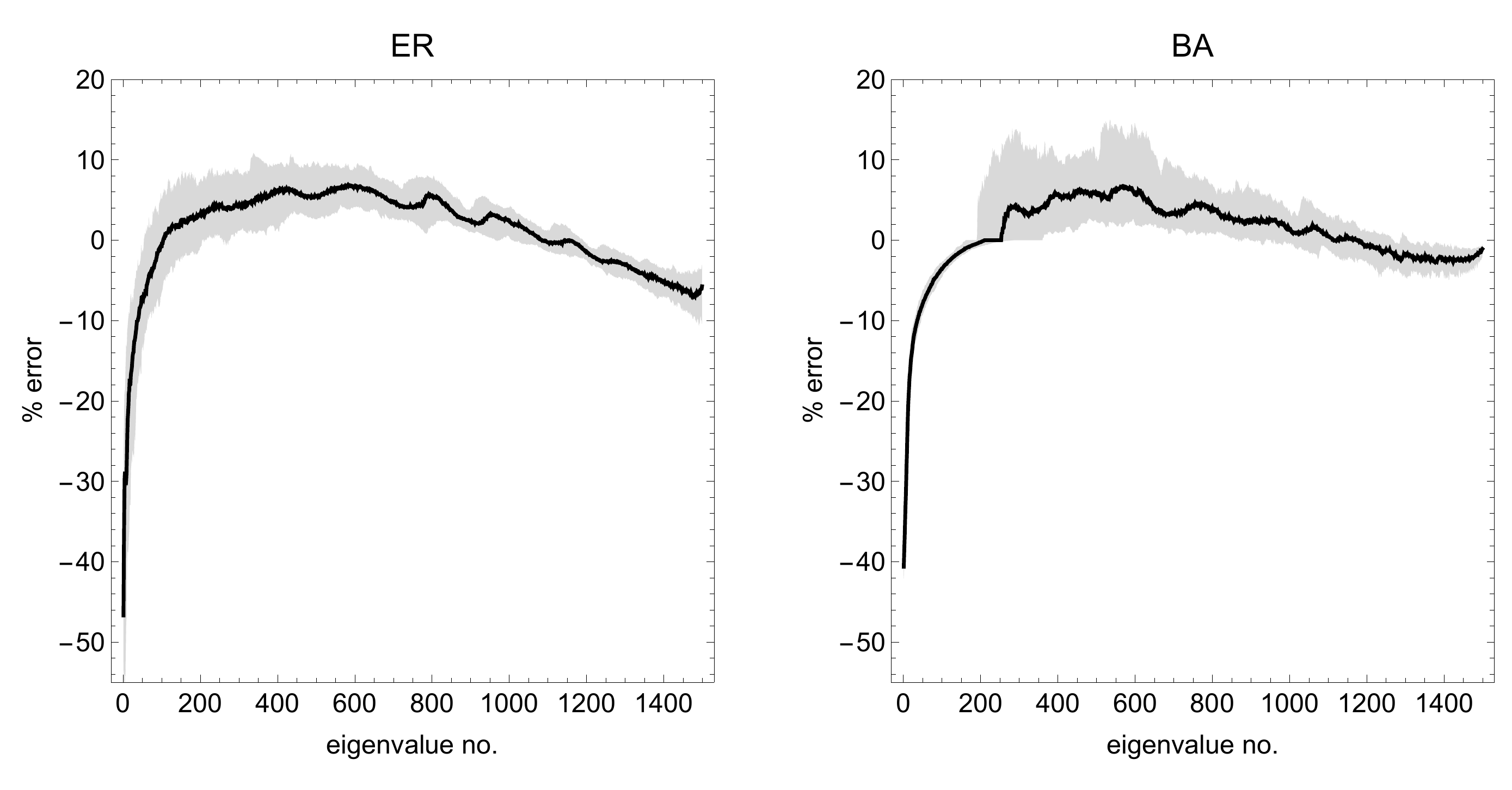}
\caption{Distributions of percentage errors in Laplacian spectra of
  direct product graphs estimated using the proposed method compared
  to actual ones. Results were collected from one hundred independent
  tests for each network topologies (``ER'' and ``BA''). Black curves
  show medians, and shaded areas show ranges from 5 to 95
  percentiles. Left (``ER''): $G$ = Erd\H os-R\'enyi random graph with
  50 nodes and 100 edges, and $H$ = Erd\H os-R\'enyi random graph with
  30 nodes and 90 edges. Right (``BA''): $G$ = Barab\'asi-Albert graph
  with 50 nodes and 2 edges per each newcomer node, and $H$ =
  Barab\'asi-Albert graph with 30 nodes and 3 edges per each newcomer
  node. Smaller eigenvalues were typically underestimated
  significantly, while estimation errors for the rest were mostly
  confined within a $\pm 10$\% range (96.5\% of the eigenvalues for
  ``ER'' and 94.0\% of the eigenvalues for ``BA'' are within this
  error range). This observation remains similar when the sizes of the
  graphs are varied (the percentage errors are smaller for larger
  graphs).}
\label{fig:performance-evaluation2-percentage-errors}
\end{figure}

Figure \ref{fig:prediction-example} shows an example of a Laplacian
spectrum of a direct product graph made of two Barab\'asi-Albert
graphs estimated using the proposed method. While there are some
noticeable differences between the actual and estimated results, the
overall profile of the spectrum is reasonably well captured. Figure
\ref{fig:performance-evaluation2-percentage-errors} shows
distributions of percentage errors of the estimated eigenvalues
compared to the actual ones. The first eigenvalue was always matched
at 0 in both actual and estimated spectra (because
Eq.~(\ref{eq:estimated-spectrum}) guarantees this), but our method
consistently underestimated the immediately following several
eigenvalues (small ones) significantly. This is where the spectrum
shows a drastic jump from 0 (see
Fig.~\ref{fig:prediction-example}). In the meantime, the estimation
errors for the remaining eigenvalues were mostly confined within a
$\pm 10$\% range for both Erd\H os-R\'enyi and Barab\'asi-Albert
cases. We consider this a reasonable estimation accuracy, given the
drastic reduction of computational complexity achieved by our
method. The characteristic shape of error profiles seen in
Fig.~\ref{fig:performance-evaluation2-percentage-errors}, i.e., a
sudden jump at the beginning followed by a gradual decrease, was
fairly consistent across various network topologies we tested, so one
might be able to develop a heuristic error reduction technique to
further improve the accuracy of estimation (which we did not attempt
within the scope of this paper).

\subsection{Estimating Laplacian spectra of strong product graphs}

Finally, we extend the methodology we used in the previous
section to estimate Laplacian spectra of strong product graphs. A
Laplacian of strong product graph $G \SP H$ is given as follows:
\begin{align}
L_{G \SP H} &= D_{G \SP H} - A_{G \SP H}\\
&=(D_{G \CP H} + D_{G \DP H}) - (A_{G \CP H} + A_{G \DP H})\\
&=(D_G \KP I_{|V_H|} + I_{|V_G|} \KP D_H + D_G \KP D_H) - (A_G \KP I_{|V_H|} + I_{|V_G|} \KP A_H + A_G \KP A_H)\\
&=D_G \KP I_{|V_H|} + I_{|V_G|} \KP D_H + D_G \KP D_H \nonumber \\
& \quad \quad \quad - (D_G - L_G) \KP I_{|V_H|} - I_{|V_G|} \KP (D_H - L_H) - (D_G - L_G) \KP (D_H - L_H)\\
&= L_G \KP I_{|V_H|} + I_{|V_G|} \KP L_H + L_G \KP D_H + D_G \KP L_H - L_G \KP L_H \label{eq:LGSPH}
\end{align}
As seen above, the only change from the direct product Laplacian
(Eq.~(\ref{eq:LGDPH})) is the inclusion of the first two terms ($L_G
\KP I_{|V_H|}$ and $I_{|V_G|} \KP L_H$), each of which has $w^G_i \KP
w^H_j$ as its eigenvector (this can be easily shown). Therefore we can
still use the same strategy to use $w^G_i \KP w^H_j$ as a reasonable
substitute of the true eigenvectors of $L_{G \SP H}$. We calculate
$L_{G \SP H} (w^G_i \KP w^H_j)$ as follows:
\begin{align}
L_{G \SP H} (w^G_i \KP w^H_j)
&= (L_G \KP I_{|V_H|} + I_{|V_G|} \KP L_H + L_G \KP D_H + D_G \KP L_H - L_G \KP L_H) (w^G_i \KP w^H_j)\\
&=\mu^G_i w^G_i \KP w^H_j + w^G_i \KP \mu^H_j w^H_j + \mu^G_i w^G_i \KP D_H w^H_j + D_G w^G_i \KP \mu^H_j w^H_j - \mu^G_i w^G_i \KP \mu^H_j w^H_j
\end{align}
Using the same aggregate notation and the heuristic assumption, this becomes
\begin{align}
L_{G \SP H} (W_G \KP W_H)
&=W_G M_G \KP W_H + W_G \KP W_H M_H + W_G M_G \KP D_H W_H + D_G W_G \KP W_H M_H - W_G M_G \KP W_H M_H\\
&\approx W_G M_G \KP W_H + W_G \KP W_H M_H + W_G M_G \KP W_H D_H + W_G D_G \KP W_H M_H - W_G M_G \KP W_H M_H\\
&= (W_G \KP W_H) \bigg(M_G + M_H + M_G \KP D_H + D_G \KP M_H - M_G \KP M_H \bigg),
\end{align}
where inside the second pair of parentheses is a diagonal matrix whose
diagonal entries show a hypothetical spectrum:
\begin{align}
&\{ \; \mu^G_i + \mu^H_j + \mu^G_i d^H_j + d^G_i \mu^H_j - \mu^G_i \mu^H_j \; \} \label{eq:estimated-spectrum2}
\end{align}

\begin{figure}[tbp]
\centering
\includegraphics[width=0.5\textwidth]{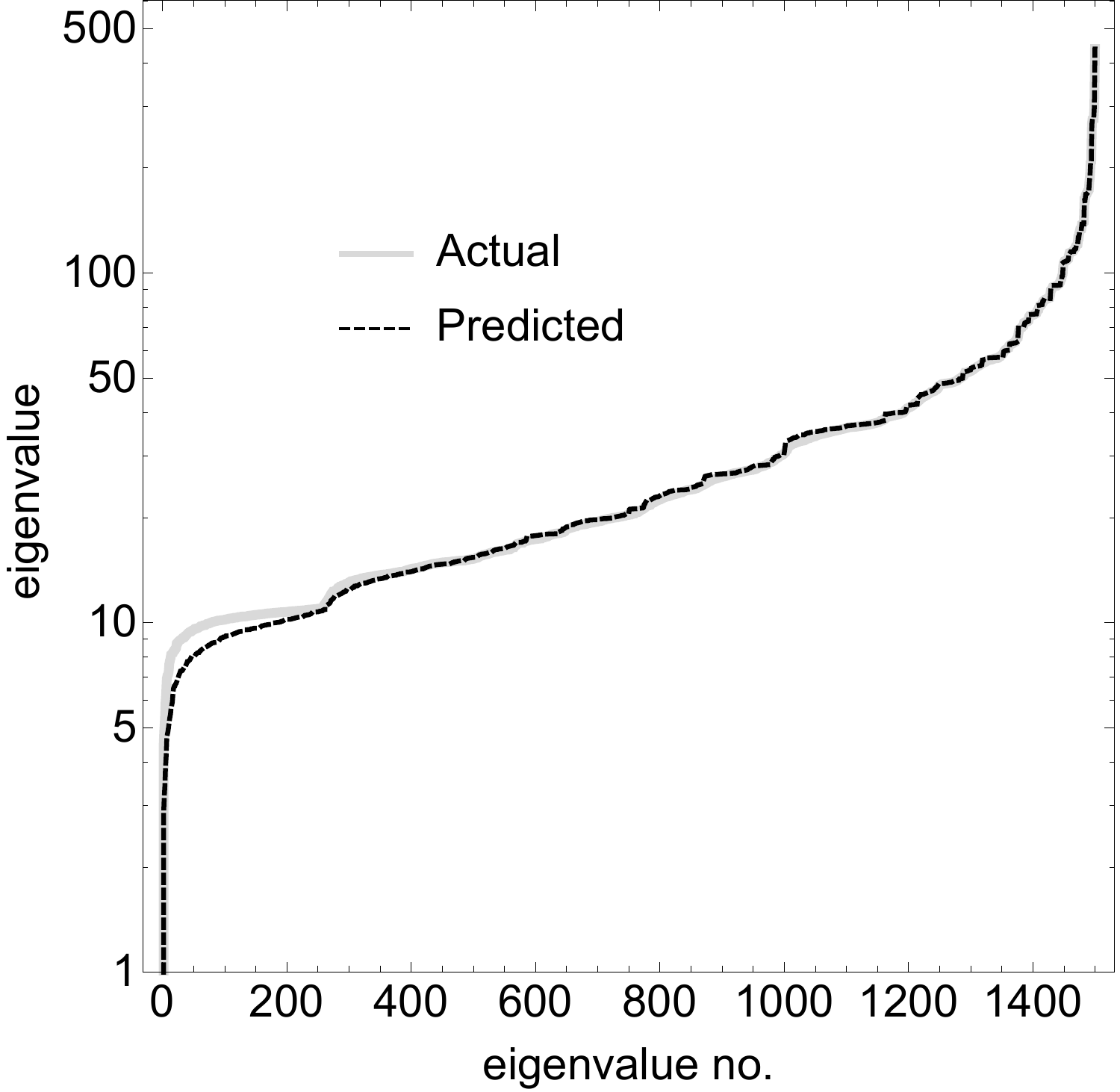}
\caption{An example of a Laplacian spectrum of a strong product graph
  estimated using the proposed method (black dashed curve) in
  comparison with the actual one (gray solid curve). Eigenvalues are
  sorted in an ascending order. $G$ = Barab\'asi-Albert graph with 50
  nodes and 2 edges per each newcomer node, and $H$ =
  Barab\'asi-Albert graph with 30 nodes and 3 edges per each newcomer
  node.}
\label{fig:prediction-example2}
\end{figure}

\begin{figure}[tbp]
\centering
\includegraphics[width=\textwidth]{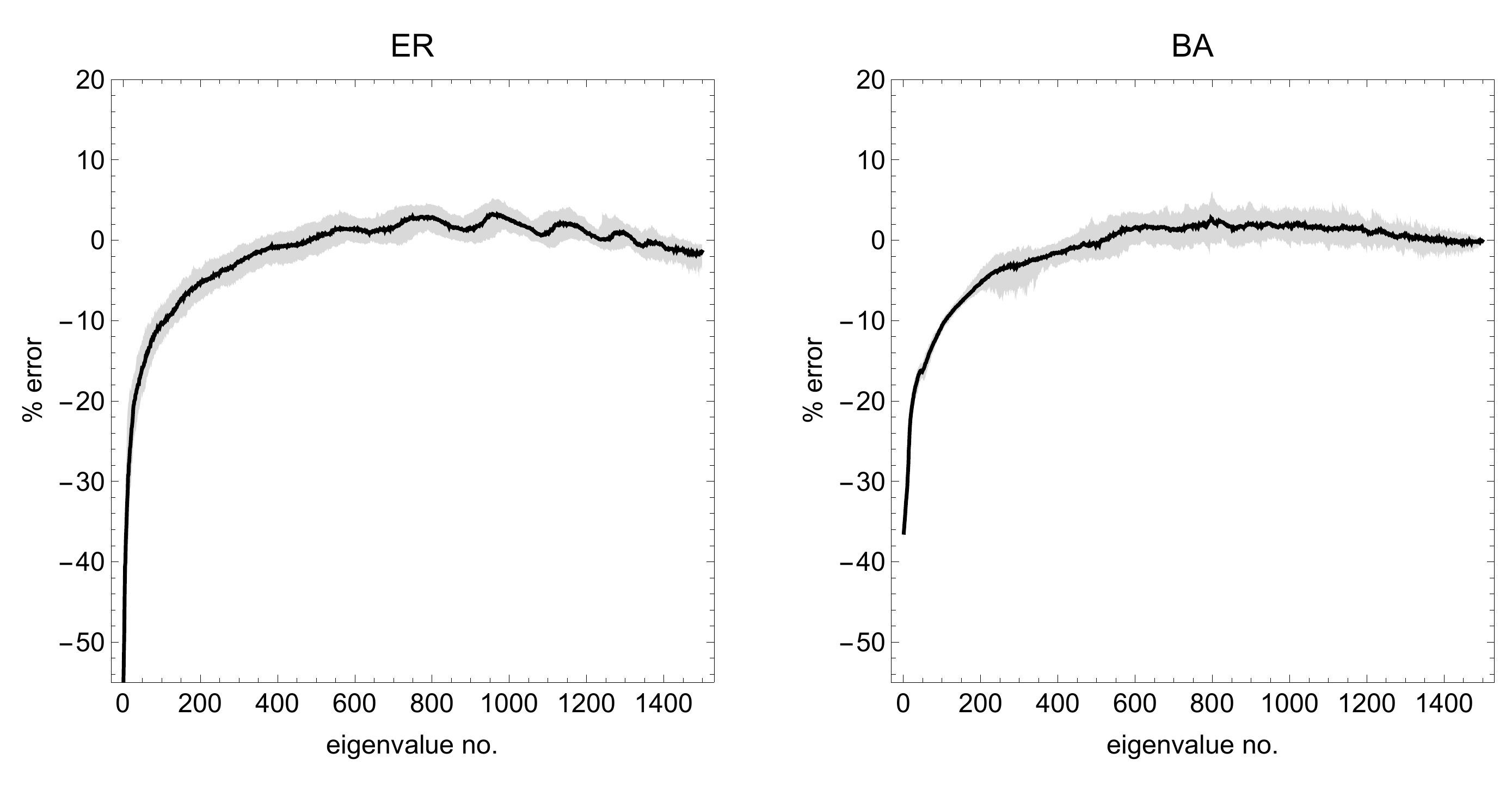}
\caption{Distributions of percentage errors in Laplacian spectra of
  strong product graphs estimated using the proposed method compared
  to actual ones. Results were collected from one hundred independent
  tests for each network topologies (``ER'' and ``BA''). Black curves
  show medians, and shaded areas show ranges from 5 to 95
  percentiles. Left (``ER''): $G$ = Erd\H os-R\'enyi random graph with
  50 nodes and 100 edges, and $H$ = Erd\H os-R\'enyi random graph with
  30 nodes and 90 edges. Right (``BA''): $G$ = Barab\'asi-Albert graph
  with 50 nodes and 2 edges per each newcomer node, and $H$ =
  Barab\'asi-Albert graph with 30 nodes and 3 edges per each newcomer
  node. Smaller eigenvalues were typically underestimated
  significantly, while estimation errors for the rest were mostly
  confined within a $\pm 10$\% range (92.9\% of the eigenvalues for
  ``ER'' and 92.8\% of the eigenvalues for ``BA'' are within this
  error range). This observation remains similar when the sizes of the
  graphs are varied (the percentage errors are smaller for larger
  graphs).}
\label{fig:performance-evaluation2-percentage-errors2}
\end{figure}

Numerical experiments showed that the correlated ordering is still
most effective in this case too (results not shown). Our proposed
method for strong product graphs can thus be summarized as follows:
\begin{framed}
\noindent {\bf Proposed method for estimating a Laplacian spectrum of
  a strong product graph}
\begin{enumerate}
\item Obtain degree and Laplacian spectra of two factor graphs $G$ and
  $H$. We denote them as $d^G=\{d^G_i \}$, $d^H=\{d^H_j\}$,
  $\mu^G=\{\mu^G_i\}$ and $\mu^H=\{\mu^H_j\}$.
\item Sort all the spectra in an ascending order.
\item Calculate the following set for all $i = 1, 2, \ldots, |V_G|$
  and $j = 1, 2, \ldots, |V_H|$:
\[ \mu^{G \SP H} = \{ \; \mu^G_i + \mu^H_j + \mu^G_i d^H_j + d^G_i \mu^H_j - \mu^G_i \mu^H_j \; \} \]
$\mu^{G \SP H}$ is the estimated Laplacian spectrum of $G \SP H$.
\end{enumerate}
\end{framed}
The computational complexity is the same as that of the method for
direct product graphs. An example of an estimated Laplacian spectrum
and the distributions of percentage errors are shown in
Figs.~\ref{fig:prediction-example2} and
\ref{fig:performance-evaluation2-percentage-errors2}, respectively.

\section{Conclusions}
\label{conclusions}

In this paper, we have developed and evaluated computationally
plausible methods for estimating Laplacian spectra of direct and
strong product graphs, for which there is no known exact formulas to
date. Our methods were designed using a few mathematically incorrect
assumptions, yet the results were generally in reasonable agreement
with the explicitly computed spectra with percentage errors confined
within a $\pm 10$\% range for most eigenvalues. The computational
complexity of the proposed methods is orders of magnitude smaller than
that of explicit computation of eigenvalues of a product graph,
especially when the factor graphs are large. This suggests that, if
one could approximate the topology of a large-scale network by a
Cartesian, direct, or strong product of two (or more) factor graphs,
the Laplacian spectrum of the network might be estimated efficiently
using our methods. Designing and evaluating such graph factorization
and spectral estimation algorithms will be an important direction of
future research.

The present study still has many fundamental limitations to which we
need to call readers' attention. The most fundamental problem is that
we have not yet come up with a rigorous mathematical explanation of
why and how the proposed methods work. We used heuristics at several
steps in designing the methods without much theoretical
support. Second, the outputs produced by our methods are nothing more
than crude estimates, which wouldn't even converge to true spectra
even if the eigenvalue orderings were completely optimized (we
confirmed this through numerical experiments conducting exhaustive
search for optimal orderings for small-sized factor graphs). The
percentage errors were particularly large for small eigenvalues, which
are often more important in spectral graph theory and network analysis
(e.g., algebraic connectivity \cite{fiedler1973algebraic}). Finally,
we used random graphs (Erd\H os-R\'enyi and Barab\'asi-Albert graphs)
in the evaluations, so there is no assurance about the behavior of our
methods on graphs with very specific non-random topologies whose
Laplacians show peculiar properties. In view of all of these
limitations, the proposed methods should be considered more as initial
``working hypotheses'' for promoting further theoretical investigation
and algorithm development, rather than immediately useful algorithms
for practical problem solving.

\section*{Acknowledgments}

The author thanks Eduardo Altmann for reviewing the draft version of
this manuscript, and the two anonymous reviewers for providing very
helpful comments. This work was conducted under financial support
offered by the Max Planck Institute for the Physics of Complex
Systems.

\bibliographystyle{plain}
\bibliography{graphproductlaplacian}

\end{document}